\documentclass[english,svgnames,dvipsnames,table]{revtex4-1}
\usepackage[T1]{fontenc}
\usepackage[latin1]{inputenc}
\usepackage{geometry}
\usepackage{float}
\usepackage{bm}
\usepackage{amsmath}
\usepackage{graphicx}
\PassOptionsToPackage{version=3}{mhchem}
\usepackage{mhchem}

\makeatletter

\providecommand{\tabularnewline}{\\}

\usepackage{xcolor}
\usepackage{tabularx}

\AtBeginDocument{
  
}

\makeatother

\usepackage{babel}
\begin{document}
\title{Nonlinear model of the firefly flash}
\author{Debashis Saikia and Madhurjya P. Bora}
\affiliation{Physics Department, Gauhati University, Guwahati 781014, India.}
\email{mpbora@gauhati.ac.in}

\begin{abstract}
A low dimensional nonlinear model based on the basic lighting mechanism
of a firefly is proposed. The basic assumption is that the firefly
lighting cycle can be thought to be a nonlinear oscillator with a
robust periodic cycle. We base our hypothesis on the well known light
producing reactions involving enzymes, common to many insect species,
including the fireflies. We compare our numerical findings with the
available experimental results which correctly predicts the reaction
rates of the underlying chemical reactions. Toward the end, a time-delay
effect is introduced for possible explanation of appearance of multiple-peak
light pulses, especially when the ambient temperature becomes low.
\end{abstract}
\maketitle

\section{Introduction}

Fireflies are fascinating insects which arouse interests in both scientists
as well as in poets! It has a very basic lighting mechanism which
is highly efficient in producing a very \emph{cold} light \cite{coldlight1}
without wasting any energy as heat, reaching an efficiency which no
man-made artificial light source has achieved so far \cite{coldlight2}.
Naturally, bioluminescence is a highly intriguing subject and a great
deal of research has been devoted to unravel the mystery of this lighting
mechanism, which have also resulted in various possible applications
\cite{applications1,applications2,applications3}.

Historically, systematic research on bioluminescence can be traced
back to as early as 1930s by Buck and Buck \cite{synchronisation2},
in which they have analysed the phenomena of large scale synchronisation
of firefly flashings. Airth and Foerster \cite{fungi1} proposed a
bioluminescence model for fungi in 1962. Stevani and Oliveira confirmed
that this proposed reaction in a live fungi using ``overlapping of
light emission spectra'' \cite{fungi2}. Several other authors also
investigated the chemical process of bioluminescence in beetles \cite{beetle1,beetle2}
and in marine snails (limpets) and luminous bacteria \cite{bacteria1}.
As far as the bioluminescence of fireflies is concerned, large scale
synchronisation of several hundreds of fireflies, usually observed
in certain geographical regions, has been subject of great deal of
attention to several researchers \cite{synchronisation1,synchronisation2,synchronisation3}.

Coming back to the biological process of firefly lighting mechanism,
it is now well known that the active material responsible for this
bioluminescence is luciferin \cite{luciferin1} and the enzyme luciferase,
which is also responsible for bioluminescence in several other species
\cite{bioluminescence1}. Despite several cutting edge researches,
unravelling the dynamics of lighting cycle of fireflies, a complete
dynamical model capable of simulating the experimental findings still
awaits us. A primary dynamical component of the firefly lighting cycle
is now known to be the so called ``nitric oxide (NO) controlled firefly
flashing'', which has been experimentally verified \cite{firfely-NO-cycle1,firfely-NO-cycle2}.
There has also been important findings related to oxygen supply required
for the firefly flash through the network of conduits in the firefly
lantern known as the \emph{tracheoles} \cite{firfely-NO-cycle3},
which has greatly contributed to our understanding of the actual lighting
mechanism though the detailed physical mechanism of the whole process
is still unclear. 

In this work, we propose a low dimensional nonlinear mathematical
model based on the basic lighting mechanism of a firefly to emulate
its flash, which we believe, is carried out for the first time. Our
basic aim is to see if the firefly lighting cycle can be thought to
be a nonlinear oscillator with a robust periodic cycle. We base our
hypothesis on the well known light producing reactions involving enzymes,
common to many insect species, including the fireflies \cite{firefly-reaction}.
An oscillator model for the firefly lighting cycle can also help us
understand the large scale synchronisation, as mentioned above. The
basic experimental input to our model is due to a series of spectroscopic
analysis carried out in-vivo and in-vitro on fireflies by several
groups \cite{firfefly-female,firfely-double-peak,firfely-male,pyrosequencing,luciferin1,firefly-spectro}.
In Section II, we construct our basic dynamical model based on the
well known set of reactions leading to the firefly lighting cycle,
where we build up a self-consistent cycle. In Section III, we carry
out a detailed bifurcation analysis of the dynamical model, proving
the existence of a robust periodic orbit with a relaxation regime.
In Section IV, we parameterise our model based on certain experimental
findings and compare our numerical results with existing data. In
Section V, we analyse the effect of time-delay on the dynamics of
the lighting cycle and possible explanation of existence of multiple-peak
light pulse, usually observed at low ambient temperature. In Section
VI, we summarise our results.

\section{The basic formalism}

\subsection{A basic mechanism}

The basic set of reactions responsible for emission of light of a
firefly can be written as \cite{firefly-reaction-1,firefly-reaction2}
\begin{eqnarray}
\ce{Luc}+\textrm{D-LH}_{2}+\ce{ATP} & \ce{<=>[\ce{Mg^{2+}}][]} & \ce{Luc.LH_{2}\textrm{-}\ce{AMP}}+\ce{PP_{i}},\label{eq:react1}\\
\ce{Luc.LH_{2}\textrm{-}\ce{AMP} +\ce{O_{2}}} & \ce{->} & \ce{Luc.Oxyluciferin^{*}}+\ce{AMP}+\ce{CO_{2}},\label{eq:react2}\\
\ce{Luc.Oxyluciferin^{*}} & \ce{->} & \ce{Luc.Oxyluciferin}+\ce{light},\label{eq:react3}
\end{eqnarray}
where luciferin $(\textrm{D-LH}_{2})$, in presence of luciferase
$(\ce{Luc})$ and adenosine triphosphate $(\ce{ATP})$ produces D-Luciferin
adenylate $(\ce{Luc.LH}_{2}\textrm{-}\ce{AMP})$. D-Luciferin adenylate
with the combination of oxygen produces oxyluciferin in the excited
state $(\ce{Oxyluciferin^{*}})$, which on de-excitation produces
the firefly light.

From the above reactions, we see that a flash is produced due to the
combination of oxygen with D-Luciferin adenylate. Denoting the concentration
of the primary active product for the bioluminescence as the D-Luciferin
adenylate as $x$ and that of oxygen as $y$, we can represent the
above reactions as
\begin{alignat}{2}
A & \ce{->} & x, & \qquad(k_{1})\label{eq:k1}\\
x+y & \ce{->} & B, & \qquad(k_{2})\label{eq:k2}
\end{alignat}
where $k_{1,2}$ are the corresponding reaction rates, $A$ are the
chain of reactants on the left of the reaction (\ref{eq:react1}),
and $B$ represents the products (\ref{eq:react2}) including oxyluciferin.
A dynamical equation based of these reaction rates can be written
now as,
\begin{equation}
\frac{dx}{dt}=k_{1}A-k_{2}xy.\label{eq:eq1}
\end{equation}

\subsection{The trigger}

\begin{figure}[t]
\begin{centering}
\includegraphics[width=0.8\textwidth]{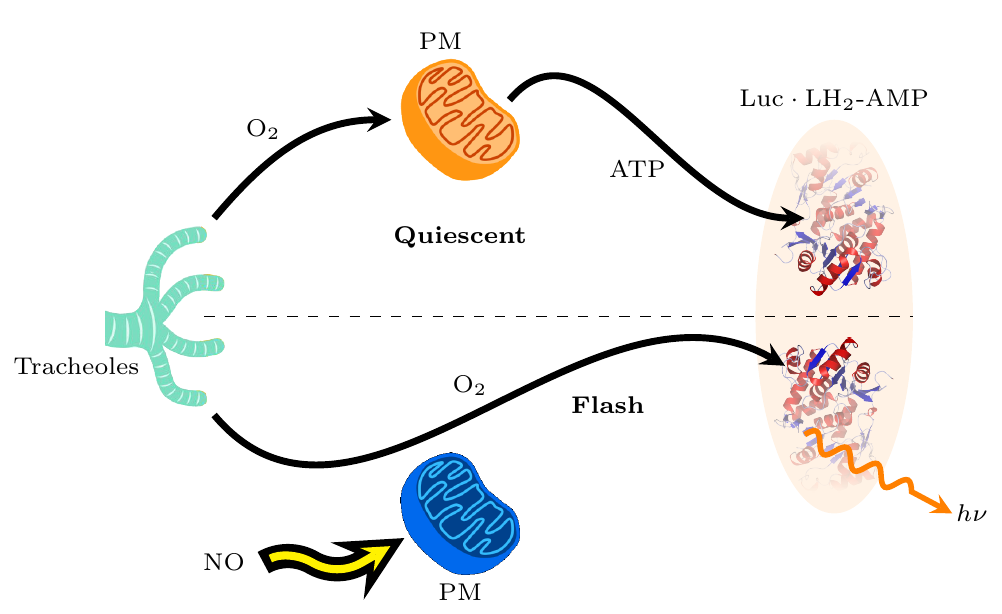}
\par\end{centering}
\caption{\label{fig:Schematic-representation-of}Schematic representation of
the NO-controlled lighting mechanism of the firefly inside its lantern.}
\end{figure}

The inhibitory property of the firefly flash suggests that there is
a kind a trigger mechanism for an emission to occur. We know that
oxygen is needed by the firefly irrespective of whether it produces
an emission or not. It is also well known that a firefly can inhibit
an emission at its will (say for example in daytime). This suppression
of emission is now known to be effected by the firefly through a controlled
release of nitric oxide (NO) \cite{firfely-NO-cycle1,firfely-NO-cycle2},
which inhibits the flow of oxygen from reacting with the D-Luciferin
adenylate, which produces oxyluciferin in the excited state $(\ce{Oxyluciferin^{*}})$.
Once the oxyluciferin in the excited state $(\ce{Oxyluciferin^{*}})$
is produced, an emission is certain. The NO-controlled mechanism of
firefly flash has been studied in quite details, which seems to explain
the lighting mechanism of the firefly flash inside its lantern, a
schematic representation of which is shown in Fig.\ref{fig:Schematic-representation-of}.
According to the NO-controlled-flash theory \cite{firfely-NO-cycle2},
a firefly lighting cycle consists of two phases --- a \emph{quiescent}
phase and a \emph{flash} phase. It is to be noted that the insect
does not have a lung and oxygen is being continuously supplied to
its body including the lantern through a network of conduits known
as the \emph{tracheoles} \cite{firfely-NO-cycle1,firfely-NO-cycle2,firfely-NO-cycle3}.
During the quiescent phase, oxygen supplied by the tracheoles binds
to the photocyte mitochondria (PM) inside the lantern, which produces
ATP through oxidative phosphorylation. The ATP then goes on to produce
the $\ce{Luc.LH}_{2}\textrm{-}\ce{AMP}$ through reaction (\ref{eq:react1})
in the peroxisomes and it keeps on accumulating inside the lantern.
Subsequently, in the flash phase, a neurological response causes octopamine
release, which activates the nitric oxide synthase (NOS) inside the
lantern, from which NO quickly diffuses and inhibits binding of oxygen
to the PMs. As a result, oxygen is made available to the peroxisomes
for interaction with the $\ce{Luc.LH}_{2}\textrm{-}\ce{AMP}$ to produce
the oxyluciferin in the excited state $(\ce{Oxyluciferin^{*}})$ and
an emission occurs.

We now hypothesise that the accumulation of D-Luciferin adenylate
$(\ce{Luc.LH}_{2}\textrm{-}\ce{AMP})$, inside the peroxisomes triggers
a neurological response which causes activation of NOS. As oxygen
is being continuously supplied by the insect through the tracheoles,
the trigger must be the critical amount of D-Luciferin adenylate that
a firefly makes available for combination with oxygen. This is equivalent
to an instability controlled by a trigger parameter, which in this
case is the critical D-Luciferin adenylate concentration $x_{c}$
and the trigger can be designed, in its simplest form, as $(x/x_{c}-1)$.
It is also experimentally known that if the amount of oxygen that
can be pumped to produce oxyluciferin, which produces the flash, is
increased, the intensity of the flash grows \cite{firefly-oxygen-effect}.
It can also be understood from the fact that if the oxygen supply
to a firefly is cut, the intensity of flash gradually decreases. Therefore,
the intensity, which should be proportional to the oxygen concentration,
in this case can be thought to be the amplitude of instability and
a dynamical relation can be written as,
\begin{equation}
\frac{dy}{dt}=\gamma\left(\frac{x}{x_{c}}-1\right)y,\label{eq:eq2-1}
\end{equation}
where $\gamma$ is the \emph{growth rate }of the equivalent linear
instability. We also add a term to take care of some amount of diffusion
of oxygen to the peroxisomes assuming that some amount of D-Luciferin
adenylate always combines with the available oxygen to produce oxyluciferin.
With this, Eq.(\ref{eq:eq2-1}) can be written as
\begin{equation}
\frac{dy}{dt}=\gamma\left(\frac{x}{x_{c}}-1\right)y+\alpha x,\label{eq:eq2}
\end{equation}
where $\alpha$ is a positive constant. In the above model, the constants
$k_{1,2}$ and $\gamma$ are positive constants.

\subsection{A feedback mechanism and delay}

The minimal dynamical model presented by Eqs.(\ref{eq:eq1},\ref{eq:eq2})
represents the basic lighting mechanism of a firefly flash, considering
one lighting cycle of a firefly. However, we note that the model lacks
a basic feedback mechanism, which must be introduced before we can
proceed. Consider the reactions given by the reactions (\ref{eq:react1}-\ref{eq:react3}).
We see that after de-excitation, oxyluciferin \emph{must} be converted
back to luciferin and then to D-Luciferin adenylate or else oxyluciferin
will be continued to be accumulated, which surely does not happen.
The details of this conversion process, however is not very well known
\cite{luciferin-conversion}. Also note the reversible nature of the
first reaction and both these observations point to the fact that
the source term i.e.\ the first term on the right of Eq.(\ref{eq:eq1})
\emph{must }be a function of $x$, $A\equiv A(x)$, so that the full
dynamical equations governing one lighting cycle of a firefly can
be written as,
\begin{eqnarray}
\frac{dx}{dt} & = & k_{1}A(x)-k_{2}xy,\label{eq:eq1-1}\\
\frac{dy}{dt} & = & \gamma\left(\frac{x}{x_{c}}-1\right)y+\alpha x,\label{eq:2-1}
\end{eqnarray}
where $A(x)$ is an yet unspecified function of $x$, which can very
well be a nonlinear function. We can now understand how this model
gives rise to oscillations. Assuming that initially $dx/dt>0$ and
$x>x_{c}$, which causes $dy/dt>0$ and $y$ increases. This will
eventually make $dx/dt<0$, provided the rate of increase of $k_{1}A(x)$
is less than that of $k_{2}xy$. The variable $y$ continues to increase
till $x=x_{c}$, which triggers the emission by making $dy/dt<0$
which causes $y$ to decrease, making $dx/dt>0$ again and the cycle
repeats. This model contains all the essential ingredients for an
emission cycle except that we still have to come up with a functional
form of $A(x)$.

It is well known that \emph{time delay} is mostly prevalent in biological
systems simply because of the reason that biological systems are \emph{not
}connected through mechanisms which can instantly transfer stimuli
(or signals) \cite{biological-delay1,biological-delay2}. We have
chosen to introduce time delay in the variable $x$ as
\begin{equation}
\frac{dx}{dt}=k_{1}A(x,x_{\tau_{1}})-k_{2}x_{\tau_{2}}y,
\end{equation}
where $x_{\tau_{1,2}}$ are the delayed values of $x\equiv x(t)$
at $t=\tau_{1,2}$, $x_{\tau_{1,2}}\equiv x(t-\tau_{1,2})$. We note
that there might be several other places including the neurological
response system of the firefly where a time delay might occur, the
details are however beyond the scope of this work. The recycling of
luciferin after de-excitation of oxyluciferin through the term $A(x)$
suggests that there can be a possible delay in that feedback mechanism,
which explains the introduction of delay in $A(x)$. The introduction
of the second delay term can be explained through delay in combination
of D-Luciferin adenylate with oxygen. For an oscillation cycle to
occur, we see that the first term in Eq.(\ref{eq:eq1-1}) \emph{must}
have a stronger dependence on $x$ than the second term if it has
to take over the second term to make $dx/dt$ positive so long as
the variable $y$ does not increase much. This indicates that $A(x)$
should obey a power law in its simplest form.

\section{Bifurcations and limit cycle --- \emph{no delay}}

In this section, we do not consider the time delay and look forward
to construct a limit cycle oscillation for our model. Mathematically,
a stable limit cycle is an isolated closed orbit of a nonlinear system
in the 2-D phase space, which attracts the nearby trajectories \cite{strogatz-book}.
Physically, the existence of a stable limit cycle points to a robust
underlying fundamental mechanism, which gives rise to the periodic
oscillation.

To normalize Eqs.(\ref{eq:eq1-1},\ref{eq:2-1}), we note that the
dimensions of the rate constants $k_{1,2}$, $[k_{1}]=T^{-1}$, which
shows that $[A(x)]=[x]$ and $[y]=1$. Besides, the dimension of the
growth rate of the trigger mechanism $[\gamma]\equiv[\alpha x]=T^{-1}$.
These provide us with the following natural scaling,
\begin{align*}
t & \to tk_{1}, & x & \to xx_{c}, &  & A(x)\to x_{c}A(x),\\
\rho & =\frac{\gamma}{k_{1}}, & \eta & =\frac{\alpha x_{c}}{k_{1}}, &  & \beta=\frac{k_{2}}{k_{1}},
\end{align*}
where we have introduced the dimensionless variables $\rho,\eta,$
and $\beta$, which represent the growth rate of the trigger instability,
amount of spontaneous emission, and ratio of the two competing reaction
rates. We note that the dimension of $A(x)$ does not prevent it from
being a nonlinear function of $x$. With these scaling and variables,
we can now write the dynamical model as,
\begin{eqnarray}
\frac{dx}{dt} & = & A(x)-\beta xy,\label{eq:dyn1}\\
\frac{dy}{dt} & = & \rho\left(x-1\right)y+\eta x,\label{eq:dyn2}
\end{eqnarray}
with the constraints $\beta,\rho,\eta>0$.

\subsection{Linear response}

\begin{figure}[t]
\begin{centering}
\includegraphics[width=0.5\textwidth]{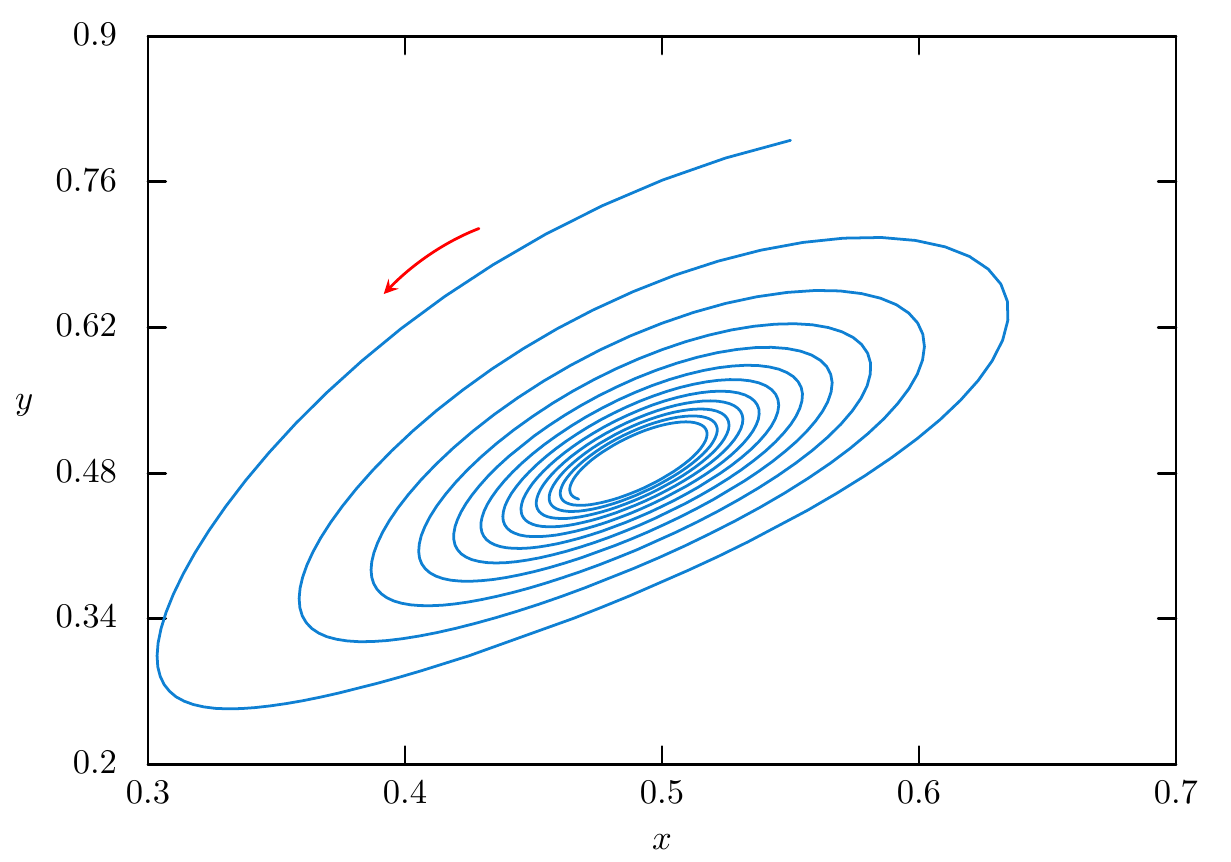}\hfill{}\includegraphics[width=0.5\textwidth]{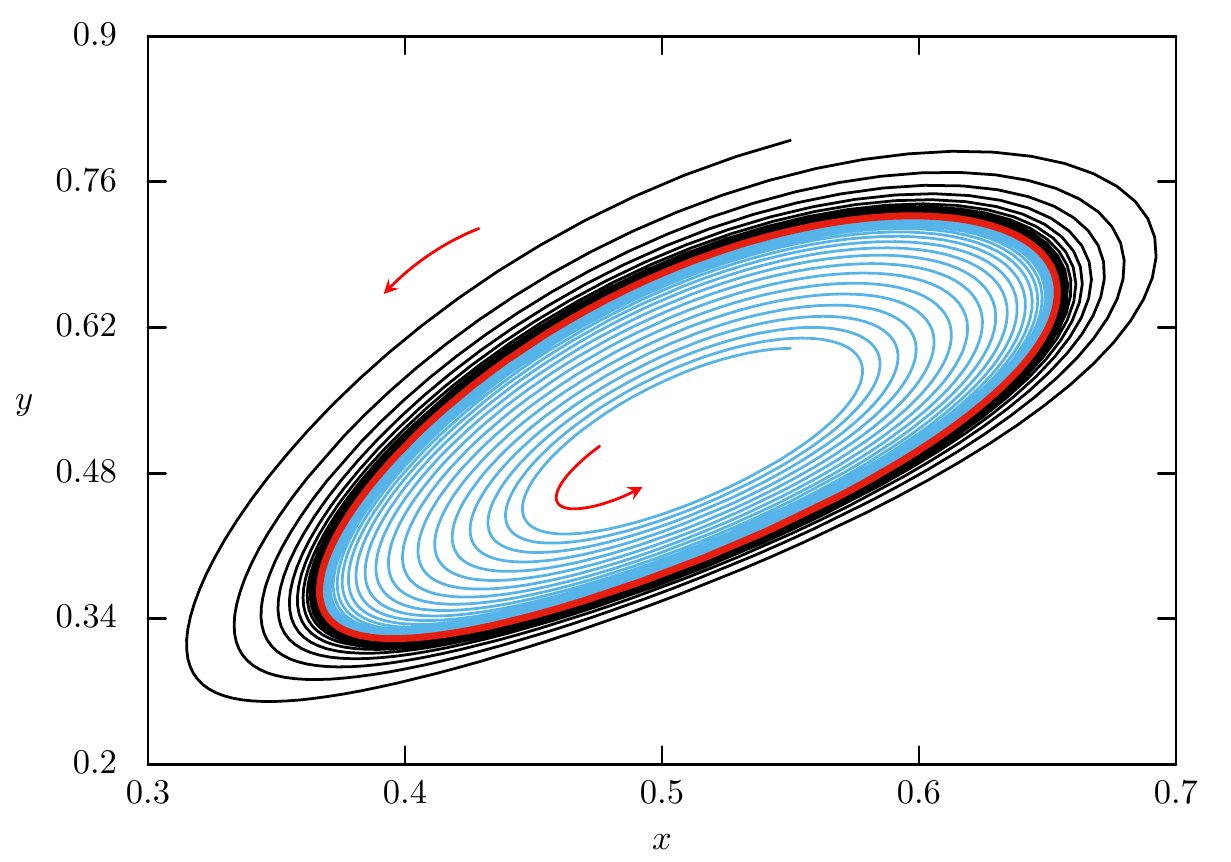}
\par\end{centering}
\caption{\label{fig:Phase-portrait-of}Phase portrait of the system for $\rho=\beta=1$.
The left panel shows the behaviour, which is an inward spiral (a sink)
above the Hopf bifurcation point, $\eta=0.51$, which is $>\eta_{{\rm cr}}$.
The right panel shows the phase portrait after Hopf bifurcation occurs
at $\eta_{{\rm cr}}$, for $\eta=0.49$. The thick red colour line
shows the resultant limit cycle.}
\end{figure}

Here, we carry out a linear analysis of the system given by Eqs.(\ref{eq:dyn1},\ref{eq:dyn2}).
In order to carry out the analysis, without loss of any generality,
we assume a simple power law function for $A(x)=x^{n}$ as described
in the before, where $n$ is a positive integer. As mentioned earlier,
for an oscillation cycle to occur, the first term on the right hand
side of Eq.(\ref{eq:dyn1}) must have a stronger dependence on $x$
than the second term, we set $n=2$ throughout our analysis. The equations
have now two sets of equilibrium points $(x^{\star},y^{\star})$ for
Eqs.(\ref{eq:dyn1},\ref{eq:dyn2}),
\begin{eqnarray}
(x^{\star},y^{\star}) & = & (0,0),\\
 & = & \left(1-\frac{\zeta}{\rho},\frac{1}{\beta}-\frac{\eta}{\rho}\right),\label{eq:eqlb-2}
\end{eqnarray}
where $\zeta=\beta\eta$. As both $x^{\star},y^{\star}>0$, the only
equilibrium point of interest is the second one. Linearsing Eqs.(\ref{eq:dyn1},\ref{eq:dyn2})
around the second equilibrium point, the Jacobian can be obtained
as \cite{strogatz-book},
\begin{equation}
J=\left(\begin{array}{cc}
1-{\displaystyle \frac{\zeta}{\rho}} & -\beta\left(1-{\displaystyle \frac{\zeta}{\rho}}\right)\\
{\displaystyle \frac{\rho}{\beta}} & -\zeta
\end{array}\right),
\end{equation}
from which we find the trace $(\tau)$ and determinant $(\Delta)$
of the Jacobian as,
\begin{eqnarray}
\tau & = & 1-\zeta\left(1+\frac{1}{\rho}\right),\\
\Delta & = & \frac{1}{\rho}(\zeta-\rho)^{2},
\end{eqnarray}
from which we can immediately see that $\Delta>0$ always and positivity
of $\tau$ depends on the condition,
\begin{equation}
\eta<\eta_{{\rm cr}}=\frac{\rho}{\beta(1+\rho)},\label{eq:crit1}
\end{equation}
with the limiting value $\eta_{{\rm lim}}\to1/\beta$ as $\rho\to\infty$.
Equivalently, the above condition can also be expressed in terms of
a critical $\rho$ value,
\begin{equation}
\rho>\rho_{{\rm cr}}=\frac{\zeta}{1-\zeta},\quad\zeta=\beta\eta<1.\label{eq:crit2}
\end{equation}
This points to the fact that we have a Hopf bifurcation occurring
at $\eta=\eta_{{\rm cr}}$. We have shown the phase portrait of the
system in Fig.\ref{fig:Phase-portrait-of}. The domain where a Hopf
bifurcation is allowed is shown in the first panel of Fig.\ref{fig:Allowed domain}.
Note that for $n=1$, $\tau=-\zeta\rho/(\zeta+\rho)$, which is always
$<0$, so a Hopf bifurcation can never exist and the possibility of
a limit cycle can be safely ruled out. 

From the inspection of the phase portrait, we conclude that a \emph{supercritical
}Hopf bifurcation occurs at $\eta=\eta_{{\rm cr}}$, following which
a limit cycle appears (see the next section). The eigenvalues of the
Jacobian are given by $\lambda=\gamma\pm i\omega$, with
\begin{eqnarray}
\gamma & = & \frac{1}{2\rho}[\rho-\zeta(1+\rho)],\\
\omega & = & \frac{1}{2\rho}\left[\rho^{2}(1-4\rho)-2\zeta\rho(1-3\rho)+\zeta^{2}(1-\rho)^{2}\right]^{1/2},
\end{eqnarray}
so that the time period $T$ of the limit cycle in the vicinity of
$\eta\simeq\eta_{{\rm cr}}$ is given by,
\begin{equation}
T=\frac{2\pi}{\omega}+{\cal O}(\eta-\eta_{{\rm cr}}).\label{eq:period}
\end{equation}
For the parameter $\beta=\rho=1,\eta=0.49$, the period of the limit
cycle is found to be $\simeq13$, which is very close to the predicted
value $T=12.33$ calculated from Eq.(\ref{eq:period}).

\subsection{Invariant region}

\begin{figure}[t]
\begin{centering}
\includegraphics[width=0.5\textwidth]{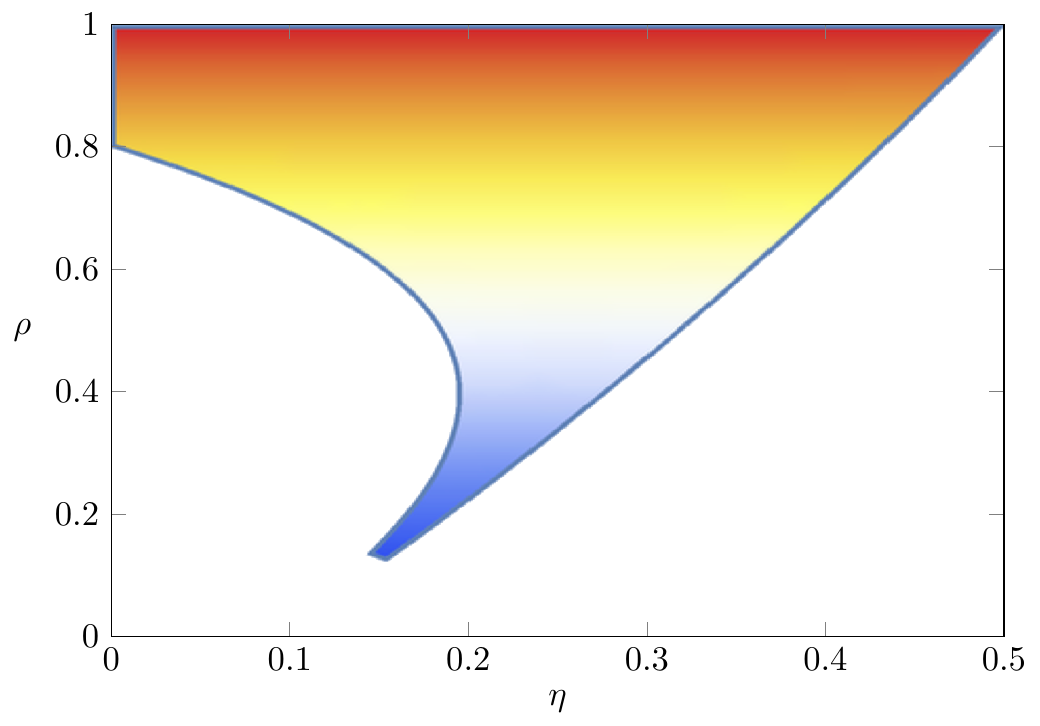}\hfill{}\includegraphics[width=0.5\textwidth]{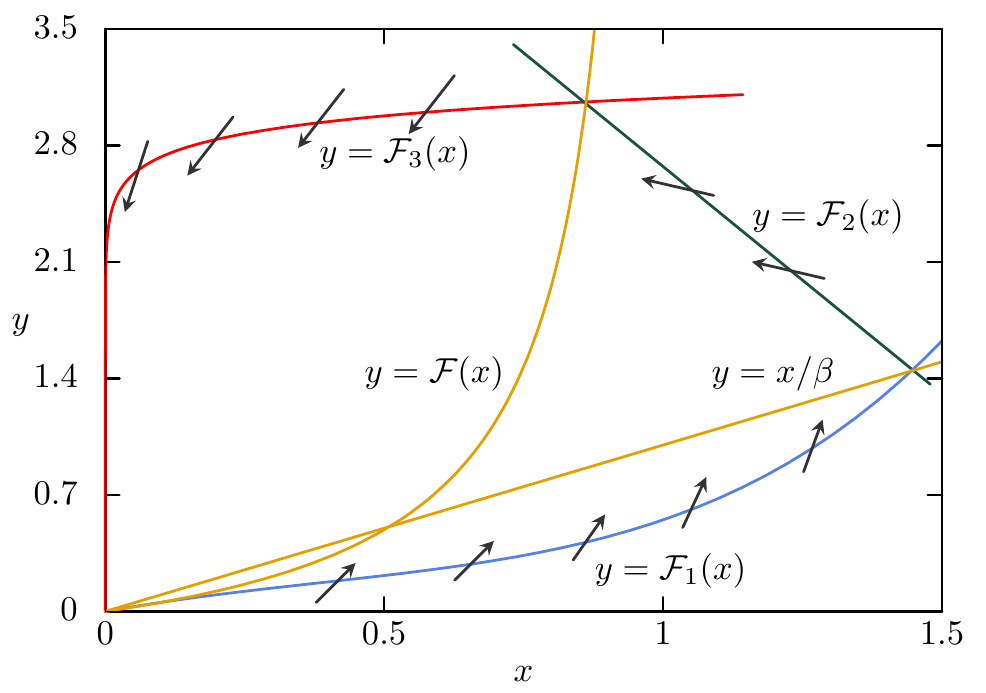}
\par\end{centering}
\caption{\label{fig:Allowed domain}Allowed domain for Hopf bifurcation in
the $(\rho,\eta)$ plane. The parameters are same as in Fig.\ref{fig:Phase-portrait-of}.
A positive invariant region for Eqs.(\ref{eq:dyn1},\ref{eq:dyn2})
around its equilibrium point. }
\end{figure}

Assuming $A(x)=x^{2}$, we now construct a \emph{positive invariant
region} around the equilibrium point of the system represented by
Eqs.(\ref{eq:dyn1},\ref{eq:dyn2}). Consider the figure shown in
the second panel of Fig.\ref{fig:Allowed domain}. The system has
three null-clines, two of which are shown in the figure. 
\begin{eqnarray}
x & = & 0,\\
y & = & x/\beta,\\
y & = & \frac{\eta x}{\rho(1-x)}={\cal F}(x).
\end{eqnarray}
 and
\begin{equation}
\frac{dy}{dx}=\frac{\rho(x-1)y+\eta x}{x^{2}-\beta xy}.\label{eq:dydx}
\end{equation}
Consider now a family of curves $y={\cal F}_{i}(x)$, given by
\begin{eqnarray}
{\cal F}_{1} & = & \frac{ax}{1+x}+bx^{4},\\
{\cal F}_{2} & = & -mx+c,\\
{\cal F}_{3} & = & \frac{dx^{p}}{1+x^{p}},
\end{eqnarray}
with certain chosen parameters $a,b,m,c,d,p>0$, such that 
\begin{eqnarray}
a,b & < & {\cal M},\\
m,d & > & {\cal N},\\
p & < & 1,
\end{eqnarray}
where ${\cal M}$ and ${\cal N}$ are two arbitrary large positive
numbers. It can now be shown that in all the regions around the equilibrium
point, the flow is always anti-clockwise \emph{into} the region, spanned
by these three curves, so that a positively invariant region can be
constructed around the equilibrium point. The region is schematically
shown in Fig.\ref{fig:Allowed domain}. The Poincar\'e-Bendixson
theorem \cite{strogatz-book,Holmes-book} now tells us that the invariant
region contains a unique limit cycle.

\section{Parameterisation and comparison --- \emph{no delay}}

\begin{figure}[t]
\begin{centering}
\includegraphics[width=0.5\textwidth]{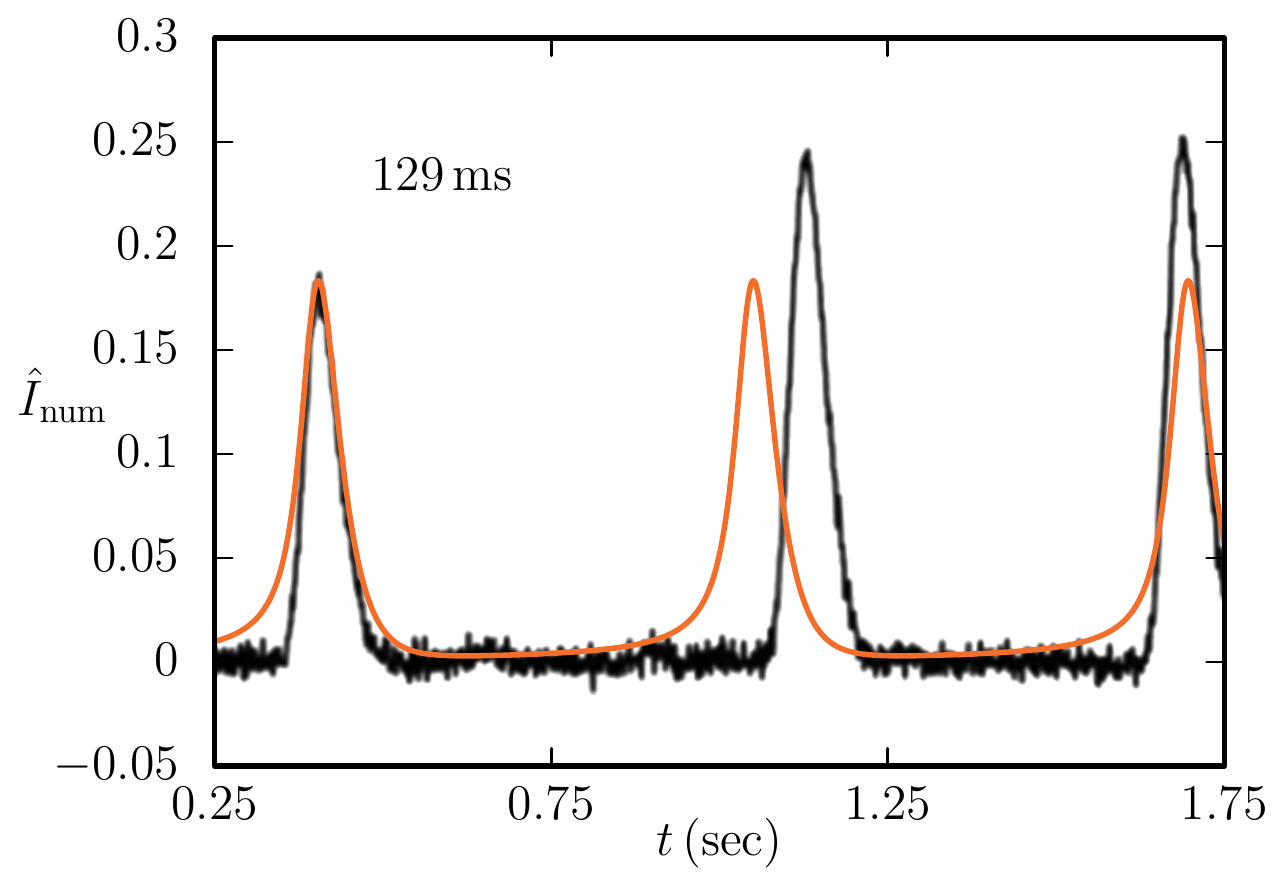}\hfill{}\includegraphics[width=0.5\textwidth]{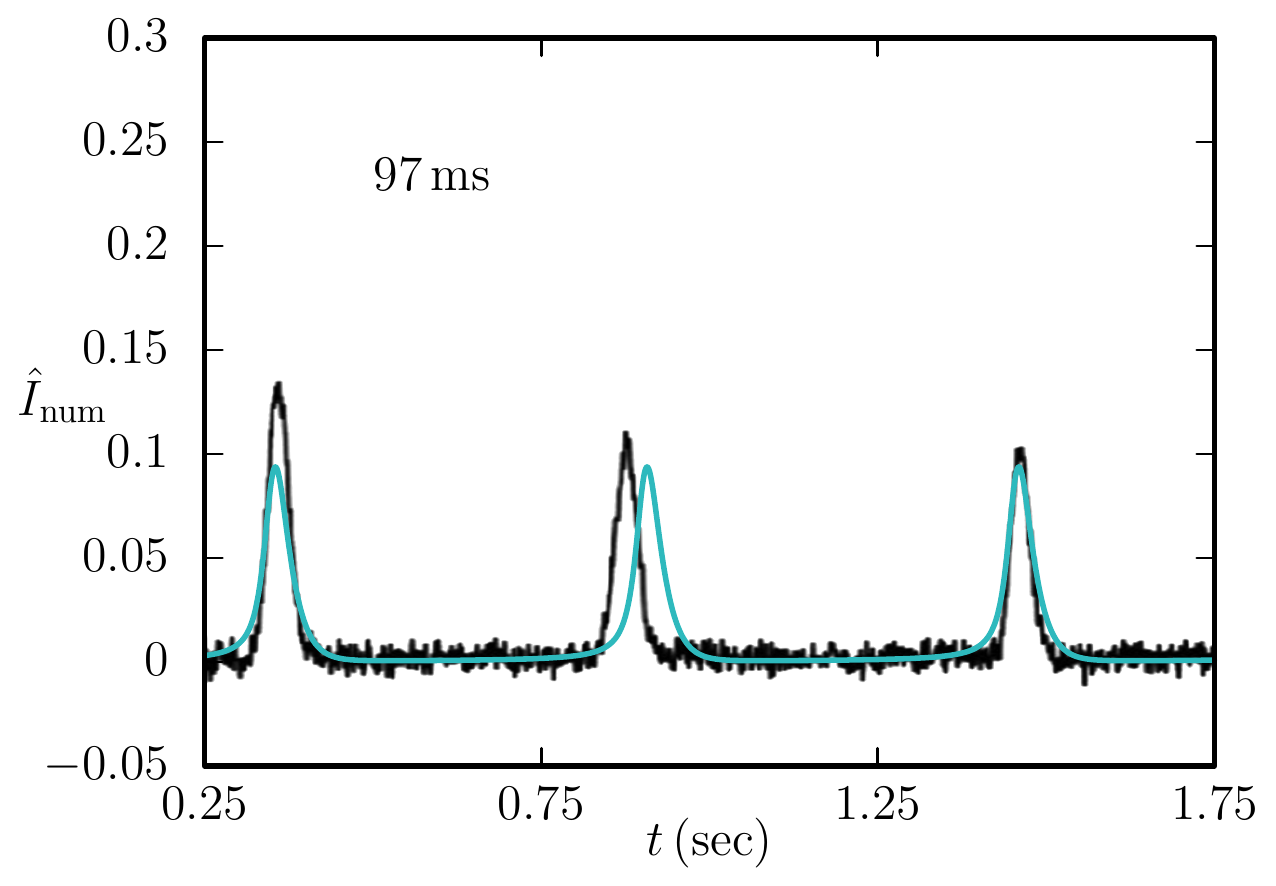}\\
\includegraphics[width=0.5\textwidth]{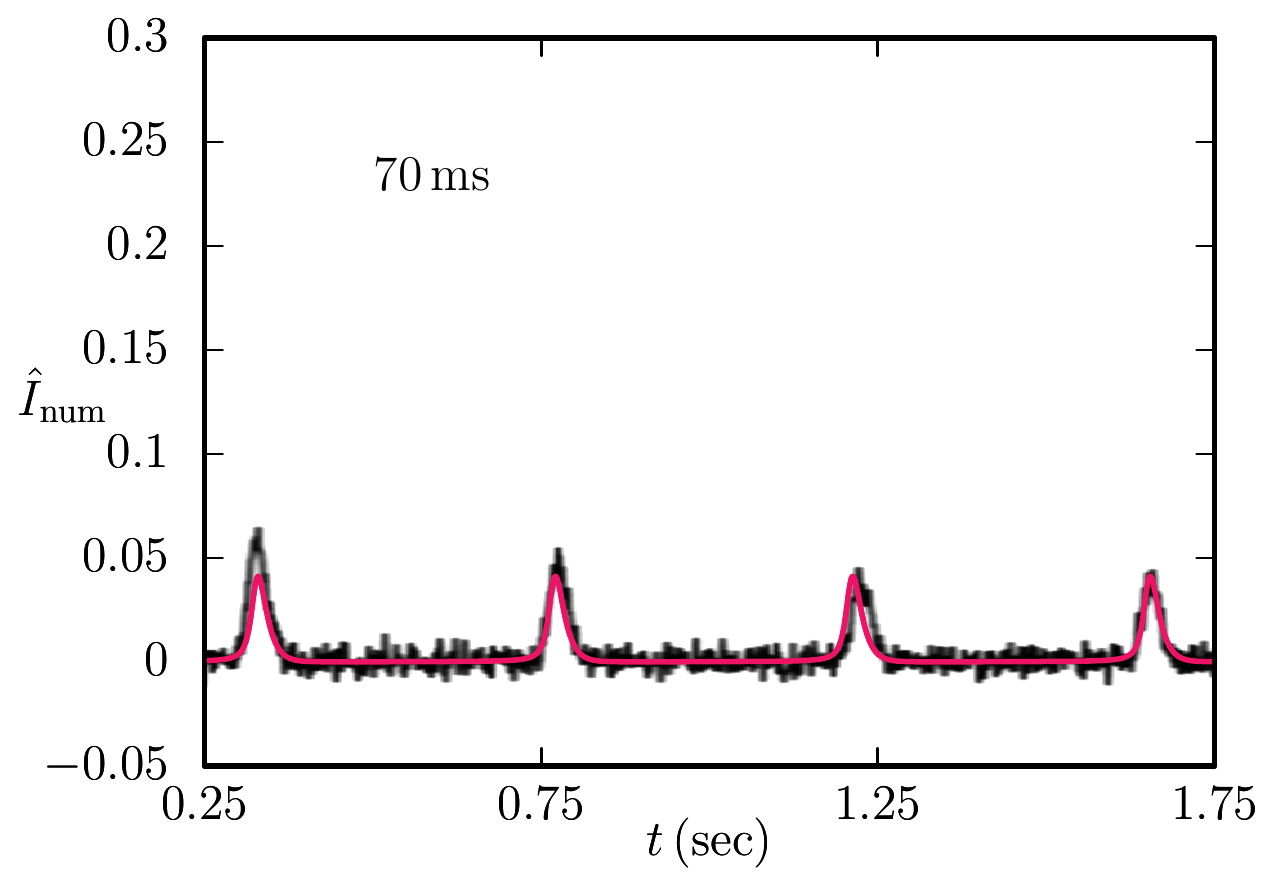}
\par\end{centering}
\caption{\label{fig:numerical results}Comparison of numerical results with
experimental data. The solid coloured lines are the results of the
numerical model, superimposed with the corresponding experimental
images \cite{firfely-male}. The values (inset) indicate the width
of the flash.}
\end{figure}

In this section, we parameterise our model for $n=2$ without any
time delay and compare the numerical results with available data.
For comparison, we take the data from the work of Sharma et al. \cite{firfely-male},
where they report about the temperature dependence on the flash duration
of male fireflies \emph{Luciola praeusta}, found in the northeastern
Indian state of Assam. In this work, the authors study the change
in the amplitude and duration on ambient temperature for a live firefly.
The actual experimental data for three ambient temperatures of $20\,{\rm ^{\circ}C}$,
$30\,{\rm ^{\circ}C}$, and $40\,{\rm ^{\circ}C}$ are shown in Fig.\ref{fig:numerical results}.
Overall, a few changes are observed --- as the ambient temperature
increases, the sharpness of the flashes are found to increase (or
a decreasing flash width), while the average amplitude of a pulse
and the duration of a lighting cycle are found to decrease. In what
follows, we try to estimate the numerical parameters on the basis
of the experimental data presented in Fig.\ref{fig:numerical results}.
There is also a companion work by the same authors on the females
of the same species \cite{firfefly-female}. We however could not
find any consistent behaviour of the flash patterns for female fireflies
with respect to their pulse width, amplitude, and period as temperature
changes in the lighting pattern of the females.

\begin{figure}[t]
\begin{centering}
\includegraphics[width=0.5\textwidth]{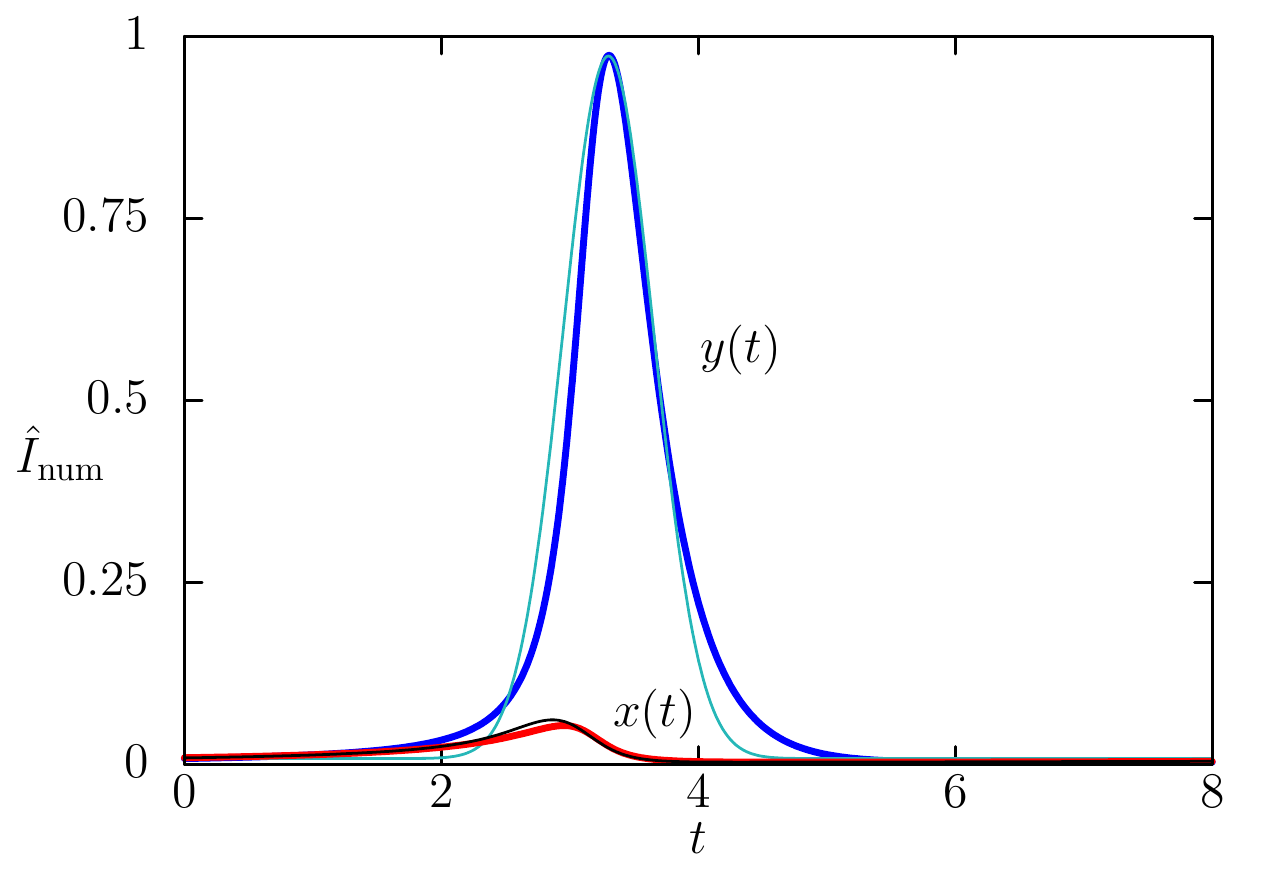}\hfill{}\includegraphics[width=0.5\textwidth]{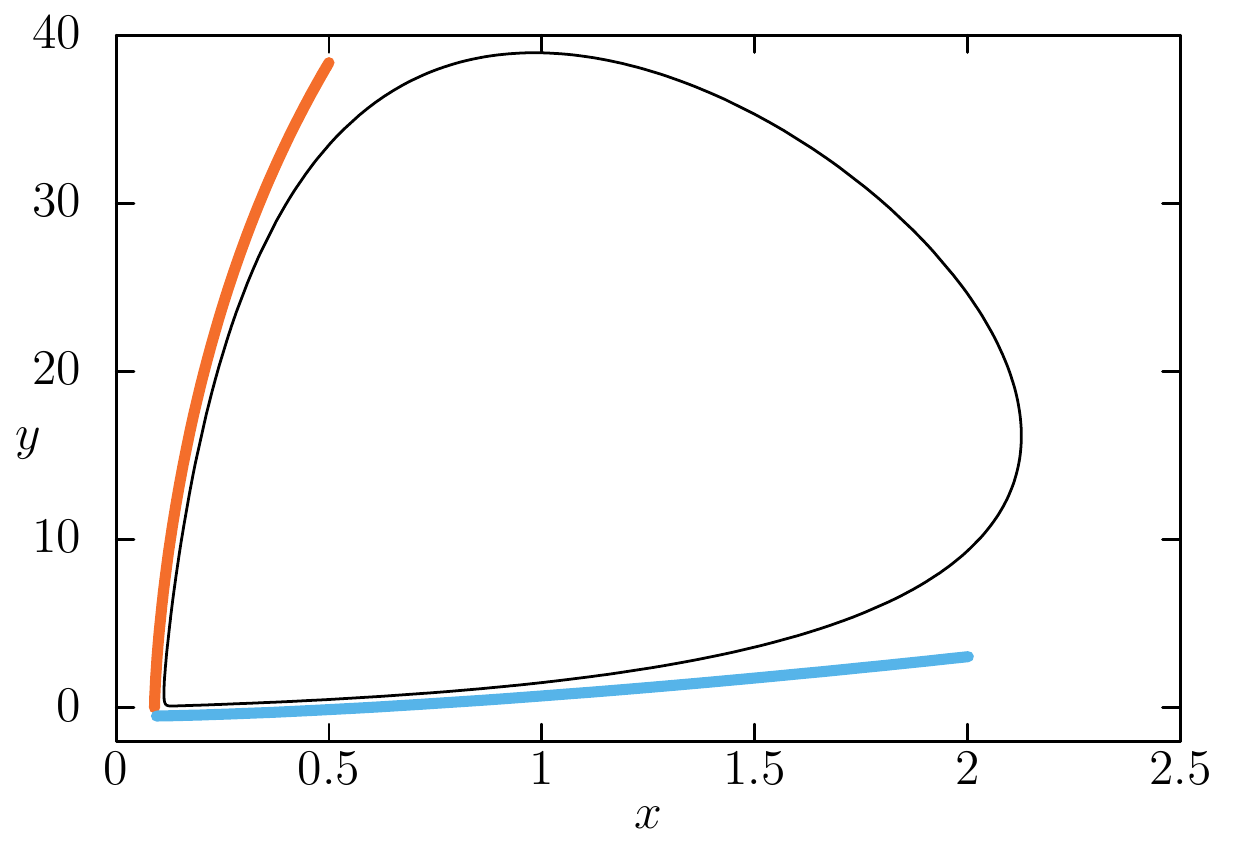}
\par\end{centering}
\caption{\label{fig:analytical approximations}The first panel shows the analytical
approximations to the numerical reproduction of the flash with the
numerical intensity normalised to unity. The slow manifold of the
lighting cycle in the $(x,y)$ phase space is shown in the second
panel.}
\end{figure}

Before we proceed, it is worth looking at the behaviour of the flash.
As can be seen from the experimental data, a lighting cycle contains
a fast and a slow manifold. Most of the time the cycle spends its
time on the slow manifold. A flash occurs on the fast manifold. In
the first panel of Fig.\ref{fig:analytical approximations}, the numerical
solution of the variable $x(t)$ and $y(t)$ are shown (blue and red
thick curves) for one flash. Note that $y(t)$ represents the intensity
of the flash, which can be approximated with an exponential function
\begin{equation}
y(t)\sim\varepsilon e^{-\mu(t-t_{0})^{2}}+\delta,\label{eq:approxy}
\end{equation}
centered around $t=t_{0}$, with the constants $\varepsilon,\mu,\delta>0$,
which is shown as the thin green curve with the best fitted parameters.
We note that the width or sharpness of the pulse is directly proportional
to $\mu$. The corresponding solution for $x(t)$ can be found in
terms of error function as,
\begin{equation}
x(t)\sim\frac{{\cal P}(t)}{{\cal C}-\int_{1}^{t}{\cal P}(\tau)\,d\tau},\quad{\cal P}(t)=\exp\left[-\delta\beta t-\frac{1}{2}\varepsilon\beta\sqrt{\frac{\pi}{\mu}}{\rm erf}\left\{ \sqrt{\mu}(t-t_{0})\right\} \right],
\end{equation}
where ${\cal C}$ is an integration constant. For the fitted parameters,
this approximation is plotted in the panel as the black thin curve
and from the figure, we see that these analytical approximations for
both $(x,y)$ work quite well during a flash. Putting back the analytical
approximation for $y(t)$ as given by Eq.(\ref{eq:approxy}) in our
dynamical model, we see that $\rho\sim2\mu$ is the parameter which
controls the sharpness of the flash or the width of the pulse ---
the larger is the value of $\rho$, the sharper is the flash or the
narrower is the width of the flash. 

Based on the above analysis, we assume that the firefly flash is a
result of a linear instability of growth rate $\gamma$, which can
be estimated from the experimental data by calculating how quickly
the the flash peaks up. If $I_{{\rm peak}}$ is the peak intensity
of a flash with a rise time (from zero) of $\Delta t_{{\rm rise}}$,
then
\begin{equation}
\frac{\Delta I}{\Delta t_{{\rm rise}}}\sim\gamma e^{\gamma t},\quad\Delta I\equiv I_{{\rm peak}},
\end{equation}
where $I\sim e^{\gamma t}$, from which $\gamma$ can be estimated
as
\begin{equation}
\gamma\sim\frac{1}{\Delta t_{{\rm rise}}}\frac{I_{{\rm peak}}}{I_{{\rm av}}}\sim\frac{2}{\Delta t_{{\rm rise}}},
\end{equation}
where $I_{{\rm av}}\equiv I_{{\rm peak}}/2$ is the average intensity
of the flash. 

In order to have an idea about period of the lighting cycle, we observe
that the duration of a lighting cycle is determined by the \emph{slow
manifold} of the cycle. In the second panel of Fig.\ref{fig:analytical approximations},
the phase portrait of a complete lighting cycle is shown. We note
that the amplitude of the variable $x(t)$ remains quite low compared
to that of $y(t)$. It can be numerically ascertained that the slow
manifold consists of a long vertical drop of the variable $y(t)$
and a near horizontal stretch of $x(t)$, which are shown in second
panel of Fig.\ref{fig:analytical approximations} as orange and blue
coloured thick lines. On the vertical manifold, we can approximate
the time as
\begin{equation}
\tau_{{\rm vert}}\approx\int_{y_{i}}^{y_{f}}\left.\frac{dt}{x^{2}-\beta xy}\right|_{x\to\beta y}\sim\frac{1}{p}\left[\ln\left(\frac{y_{i}}{y_{f}}\right)+\ln\left(\frac{p-y_{f}\beta\rho}{p-y_{i}\beta\rho}\right)\right],
\end{equation}
where $y_{i,f}$ is the initial and final values of $y(t)$ on the
vertical drop and the parameter $p=\rho-\beta\eta$. On the horizontal
manifold, the time spent in the phase space can be approximated as
\begin{eqnarray}
\tau_{{\rm hor}} & \approx & \int_{x_{i}}^{x_{f}}\left.\frac{dt}{\rho(x-1)+\eta x}\right|_{y\to\eta x\rho^{-1}/(x-1)},\nonumber \\
 & \sim & \frac{\rho}{x_{i}x_{f}p^{2}}\left[p(x_{f}-x_{i})+x_{i}x_{f}\beta\eta\left\{ \ln\left(\frac{x_{f}}{x_{i}}\right)+\ln\left(\frac{p-x_{i}\rho}{p-x_{f}\rho}\right)\right\} \right],
\end{eqnarray}
where $x_{i,f}$ is the initial and final values of $x(t)$ on the
horizontal stretch. So, the approximate time spent on the slow manifold
is
\begin{equation}
\tau_{{\rm slow}}\sim\tau_{{\rm vert}}+\tau_{{\rm hor}}\propto(\rho-\beta\eta)^{-1}.\label{eq:slow}
\end{equation}
Or, in other words, the period of a lighting cycle should be inversely
proportional to the quantity $(\rho-\beta\eta$).

With the value of $\gamma=\gamma_{{\rm expt}}$ estimated from the
experimental data as described above and with Eq.(\ref{eq:slow}),
we parameterise the model given by Eqs.(\ref{eq:dyn1},\ref{eq:dyn2})
for the free parameters $k_{1,2},\eta$, for an optimum fit, subject
to the constraints $\eta<\eta_{{\rm lim}},\gamma=\gamma_{{\rm expt}}$,
where we minimise the quantity
\begin{equation}
\epsilon=\frac{1}{2}\left(\left|\hat{I}_{{\rm num}}-\bar{I}_{{\rm peak}}\right|+\left|t_{{\rm num}}-t_{{\rm period}}\right|\right),
\end{equation}
which is the combined, average absolute fitting error with respect
to the amplitude and period of the pulse, where $t_{{\rm num}}$ is
the numerical period, $t_{{\rm period}}$ is the period found from
the experimental data, $\hat{I}_{{\rm num}}$ is the numerical amplitude
of the flash, and $\bar{I}{}_{{\rm peak}}$ experimental peak intensity.
The numerical fits to the available experimental data are shown in
Fig.\ref{fig:numerical results}., as solid colour lines. The fitted
parameters are listed in Table.\ref{tab:Parametrisation-of-numerical}.
\begin{table}[t]
\caption{\label{tab:Parametrisation-of-numerical}Parameterisation of numerical
model.}
~\\
~
\begin{centering}
\begin{tabular}{|c|c|c|c|c|c|c|c|c|c|c|c|c|c|}
\cline{1-10} \cline{2-10} \cline{3-10} \cline{4-10} \cline{5-10} \cline{6-10} \cline{7-10} \cline{8-10} \cline{9-10} \cline{10-10} \cline{12-14} \cline{13-14} \cline{14-14} 
\vrule width0pt height12pt depth8pt Temp. & $t_{{\rm period}}$  & $t_{{\rm flash}}$ & $\gamma$ & $k_{1}$ & $k_{2}$ & $\eta_{{\rm lim}}$ & $\eta$ & $\bar{I}{}_{{\rm peak}}$ & $\hat{I}_{{\rm num}}$ & \multicolumn{1}{c|}{} & $\rho$ & $\beta$ & $t_{{\rm num}}$\tabularnewline
\cline{1-10} \cline{2-10} \cline{3-10} \cline{4-10} \cline{5-10} \cline{6-10} \cline{7-10} \cline{8-10} \cline{9-10} \cline{10-10} \cline{12-14} \cline{13-14} \cline{14-14} 
\vrule width0pt height12pt depth8pt  $20\,{\rm ^{\circ}C}$ & $0.646$ & $0.129$ & $39.63$ & $22$ & $1.15$ & $12.30$ & $8.20$ & $0.183$ & $0.183$ &  & $1.80$ & $0.052$ & $14.21$\tabularnewline
\cline{1-10} \cline{2-10} \cline{3-10} \cline{4-10} \cline{5-10} \cline{6-10} \cline{7-10} \cline{8-10} \cline{9-10} \cline{10-10} \cline{12-14} \cline{13-14} \cline{14-14} 
\vrule width0pt height12pt depth8pt  $30\,{\rm ^{\circ}C}$ & $0.548$ & $0.097$ & $58.41$ & $24$ & $3.42$ & $5.03$ & $2.06$ & $0.116$ & $0.094$ &  & $2.43$ & $0.143$ & $13.15$\tabularnewline
\cline{1-10} \cline{2-10} \cline{3-10} \cline{4-10} \cline{5-10} \cline{6-10} \cline{7-10} \cline{8-10} \cline{9-10} \cline{10-10} \cline{12-14} \cline{13-14} \cline{14-14} 
\vrule width0pt height12pt depth8pt  $40\,{\rm ^{\circ}C}$ & $0.442$ & $0.070$ & $91.24$ & $27$ & $12.0$ & $1.74$ & $0.67$ & $0.051$ & $0.041$ &  & $3.38$ & $0.444$ & $11.93$\tabularnewline
\cline{1-10} \cline{2-10} \cline{3-10} \cline{4-10} \cline{5-10} \cline{6-10} \cline{7-10} \cline{8-10} \cline{9-10} \cline{10-10} \cline{12-14} \cline{13-14} \cline{14-14} 
\end{tabular}
\par\end{centering}
\end{table}
 In the table, the $\hat{I}_{{\rm num}}$ is the numerical amplitude
of the flash, normalised by the average experimental peak intensity
$\bar{I}{}_{{\rm peak}}$ for $20\,{\rm ^{\circ}C}$ with $\epsilon_{{\rm max}}=0.5$.

The best fit parameters indicate that the reaction rates $k_{1,2}$
increase with temperature, which for any reactions involving enzymes,
is expected to be true \cite{pyrosequencing}. It is also interesting
to note that the experimentally reported values for the reaction rates
$k_{1,2}$ for firefly luciferase are $19.2\,{\rm s}^{-1}$ and $0.96\,{\rm s}^{-1}$
respectively \cite{pyrosequencing}, which are of the same order as
our numerically fitted values (see Table.\ref{tab:Parametrisation-of-numerical})
for $20\,{\rm ^{\circ}C}$. In the same report, it is also reported
that the Pyrosequencing experimental data predicts $k_{1}=30\pm10\,{\rm s}^{-1}$
and $k_{2}=10\pm3\,{\rm s}^{-1}$, with a significant change in the
value of $k_{2}$ \cite{pyrosequencing}. Interestingly, the Pyrosequencing
data are of the same order for our numerically fitted values at higher
temperature. As can be seen from Fig.\ref{fig:numerical results},
our fitted results most agree with the experimental data at higher
temperatures, namely at $40\,{\rm ^{\circ}C}$ and as reported in
the paper by Sharma et al. \cite{firfely-male}, the natural flashing
temperature of these fireflies lies within $27-32\,{\rm ^{\circ}C}$,
which is the ambient temperature of the habitats where these fireflies
are found.

\section{Effect of delay}

In this section, we examine what happens to the lighting cycle, when
the combination of oxygen with D-Luciferin adenylate as well as the
conversion of oxyluciferin to luciferin through the feedback cycle
are delayed
\begin{eqnarray}
\frac{dx}{dt} & = & xx_{\tau_{1}}-\beta x_{\tau_{2}}y=f(x,x_{\tau_{1}},x_{\tau_{2}},y),\label{eq:delay-1}\\
\frac{dy}{dt} & = & \rho\left(x-1\right)y+\eta x=g(x,y),\label{eq:delay-2}
\end{eqnarray}
where $x_{\tau_{1,2}}\equiv x(t-\tau_{1,2})$ represent the value
of the variable $x$ at an earlier times $t-\tau_{1,2}$, with $\tau_{1,2}$
being the delays. The equilibrium point of interest is given by
\begin{equation}
(x^{\star},y^{\star})=\left(1-\frac{\zeta}{\rho},\frac{1}{\beta}-\frac{\eta}{\rho}\right).
\end{equation}
Following standard procedure \cite{Rand,Roussel}, we now expand the
variables about the equilibrium point,
\begin{eqnarray}
\bm{x} & = & \bm{x}^{\star}+\delta\bm{x}\\
\dot{\bm{x}} & = & \delta\dot{\bm{x}}=\bm{f}\left(\bm{x}^{\star}+\delta\bm{x},\bm{x}^{\star}+\delta\bm{x}_{1},\bm{x}^{\star}+\delta\bm{x}_{2}\right),
\end{eqnarray}
where $\delta x$ and $\delta y$ are the respective small deviations
of the variables at time $t$, with
\begin{eqnarray}
\bm{x} & = & (x,y)',\\
\delta\bm{x} & = & (\delta x,\delta y)',\\
\bm{f} & = & (f,g)',
\end{eqnarray}
where $()'$ denotes the transpose of a row vector. The deviations
$\delta\bm{x}_{1,2}$ represent the deviations at the delayed time
$t-\tau_{1,2}$. Expanding and linearising Eqs.(\ref{eq:delay-1},\ref{eq:delay-2})
about the equilibrium point, we have
\begin{equation}
\delta\dot{\bm{x}}\simeq\bm{J}_{0}\,\delta\bm{x}+\bm{J}_{1}\,\delta\bm{x}_{1}+\bm{J}_{2}\,\delta\bm{x}_{2},\label{eq:delay}
\end{equation}
where $\bm{J}_{0,1,2}$ are the linearised Jacobians with respect
to time $t$ (no delay) and delayed time $t-\tau_{1,2}$,
\begin{eqnarray}
\bm{J}_{0} & = & \left(\begin{array}{cc}
1-{\displaystyle \zeta/\rho} & -\beta+{\displaystyle \beta\zeta/\rho}\\
{\displaystyle \rho/\beta} & -\zeta
\end{array}\right),\\
\bm{J}_{1} & = & \left(\begin{array}{cc}
1-\zeta/\rho & 0\\
{\displaystyle 0} & 0
\end{array}\right),\\
\bm{J}_{2} & = & \left(\begin{array}{cc}
-1+\zeta/\rho & 0\\
{\displaystyle 0} & 0
\end{array}\right),
\end{eqnarray}
Eq.(\ref{eq:delay}) represents the linear delay differential equation
(DDE) corresponding to the DDE Eqs.(\ref{eq:delay-1},\ref{eq:delay-2}).
Assume now that the deviations $\delta\bm{x}(t)$ has an exponential
solution of the form
\begin{equation}
\delta\bm{x}(t)\sim\bm{A}e^{\lambda t}.
\end{equation}
Substituting the above trial solution to the linear DDE, we arrive
at the characteristic equation
\begin{equation}
\left|\bm{J}_{0}+e^{-\lambda\tau_{1}}\bm{J}_{1}+e^{-\lambda\tau_{2}}\bm{J}_{2}-\lambda\bm{I}\right|=0,
\end{equation}
which should possess complex solutions of the type $\lambda=\gamma\pm i\omega$.
We now look for the possibility of having a pair of pure imaginary
eigenvalues $\lambda=\pm i\omega$, giving rise to a Hopf bifurcation
\cite{Rand,Roussel}. At the Hopf point, separating the real and imaginary
parts of the the characteristic equation, we have,
\begin{eqnarray}
(\zeta-\rho)[\zeta c_{1}+\omega s_{1}]+(\zeta-\rho)^{2}-\rho\omega^{2} & = & 0,\label{eq:hopf1}\\
\omega[(\zeta-1)\rho+\zeta]-(\zeta-\rho)[\zeta s_{1}-\omega c_{1}] & = & 0,\label{eq:hopf2}
\end{eqnarray}
where 
\begin{eqnarray}
\cos(\omega\tau_{1})-\cos(\omega\tau_{2}) & = & c_{1},\\
\sin(\omega\tau_{1})-\sin(\omega\tau_{2}) & = & s_{1}.
\end{eqnarray}

\subsection{Double peak pulse}

We would now like to note the occurrence of pulse with multiple peaks,
especially the occurrence of double peak pulse at low temperature
\cite{firfely-double-peak}. Following the trends of dependence of
various parameters on temperature (see Table.\ref{tab:Parametrisation-of-numerical})
we look for a regime with temperature lower than $20^{\circ}\,{\rm C}$,
where multiple peak pulse may occur with the constraints $\rho<1.80,\beta<0.052,\eta>8.20,$
and $t_{{\rm num}}>14.21$. 
\begin{figure}[t]
\begin{centering}
\includegraphics[width=0.5\textwidth]{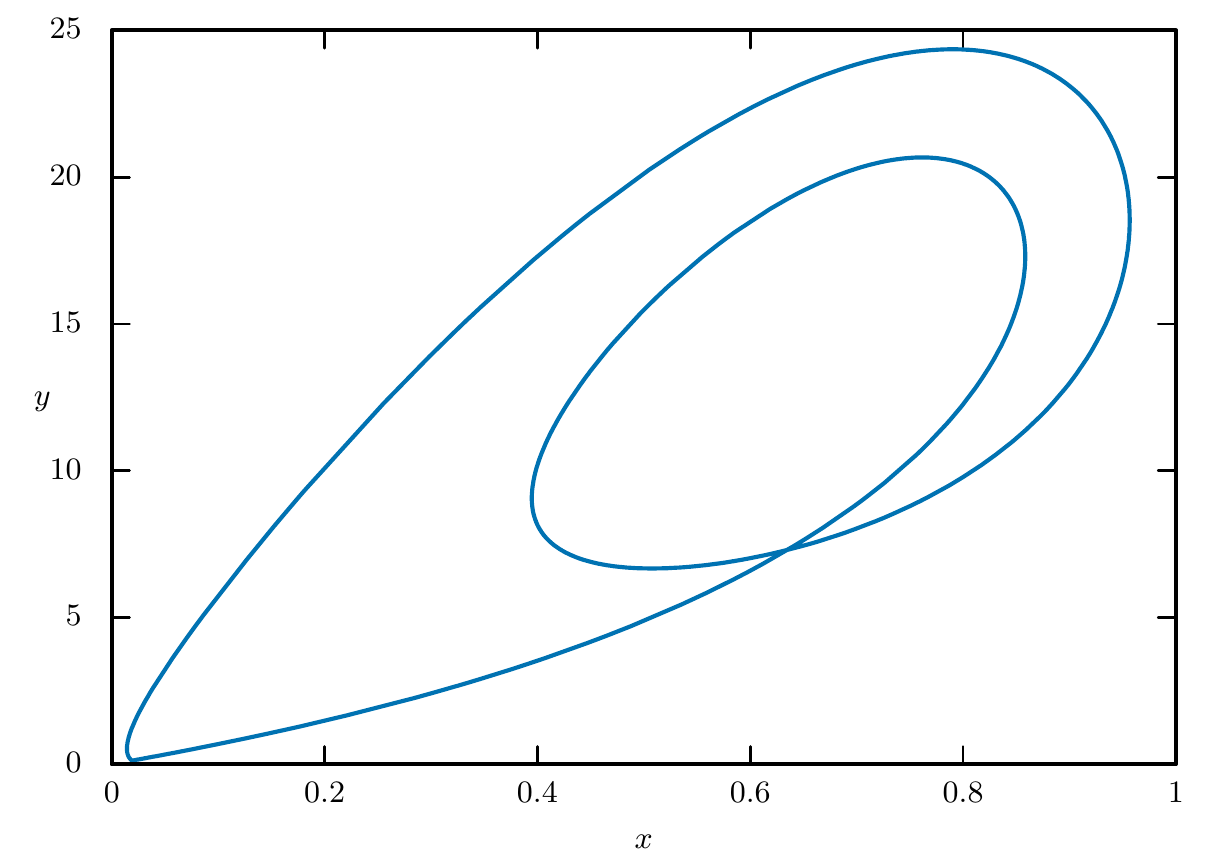}\hfill{}\includegraphics[width=0.5\textwidth]{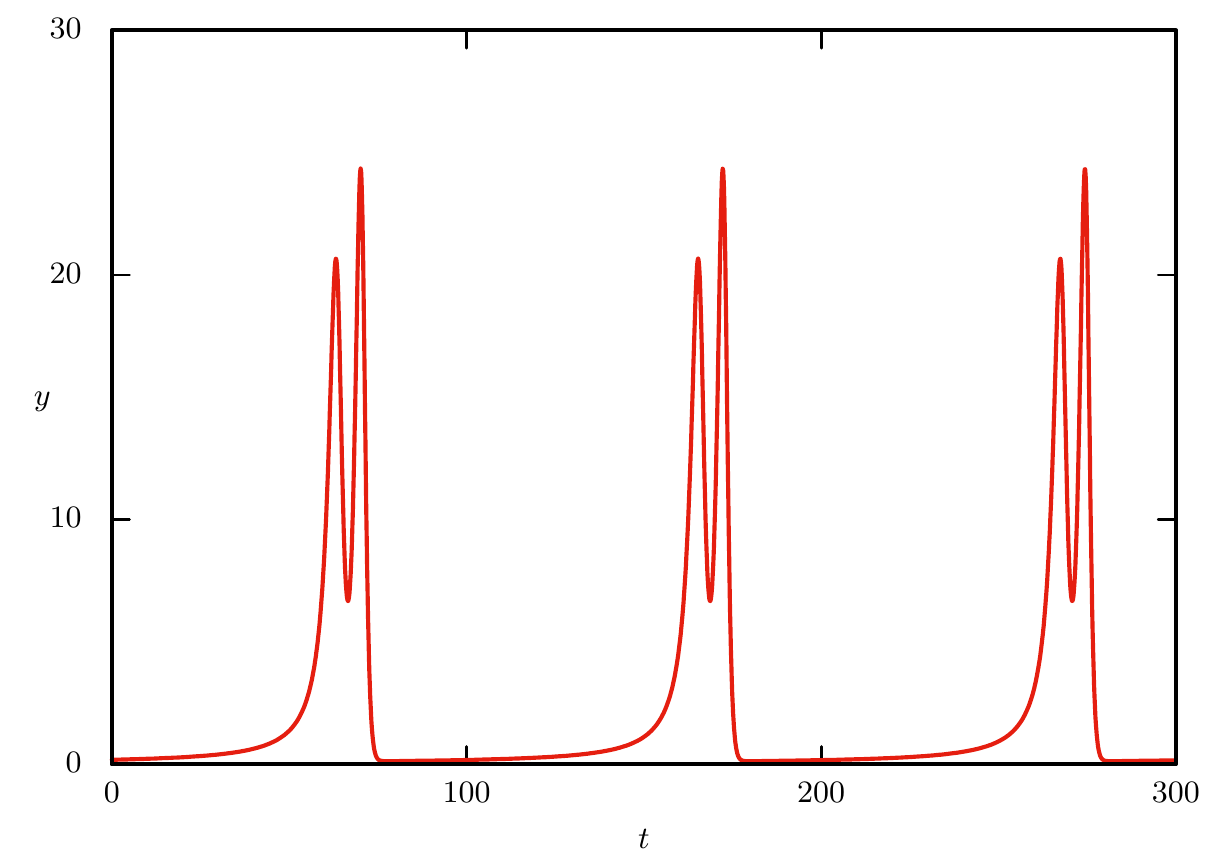}
\par\end{centering}
\caption{\label{fig:The-phase-portrait}The phase portrait of a double-peak
pulse is shown in the first panel and the corresponding pulse for
the parameters $\rho=1.53,\beta=0.045,\eta=9.9,\tau_{1}=3.4,\tau_{2}=2.805$
is shown in the second panel.}
\end{figure}

\begin{figure}[t]
\begin{centering}
\includegraphics[width=0.5\textwidth]{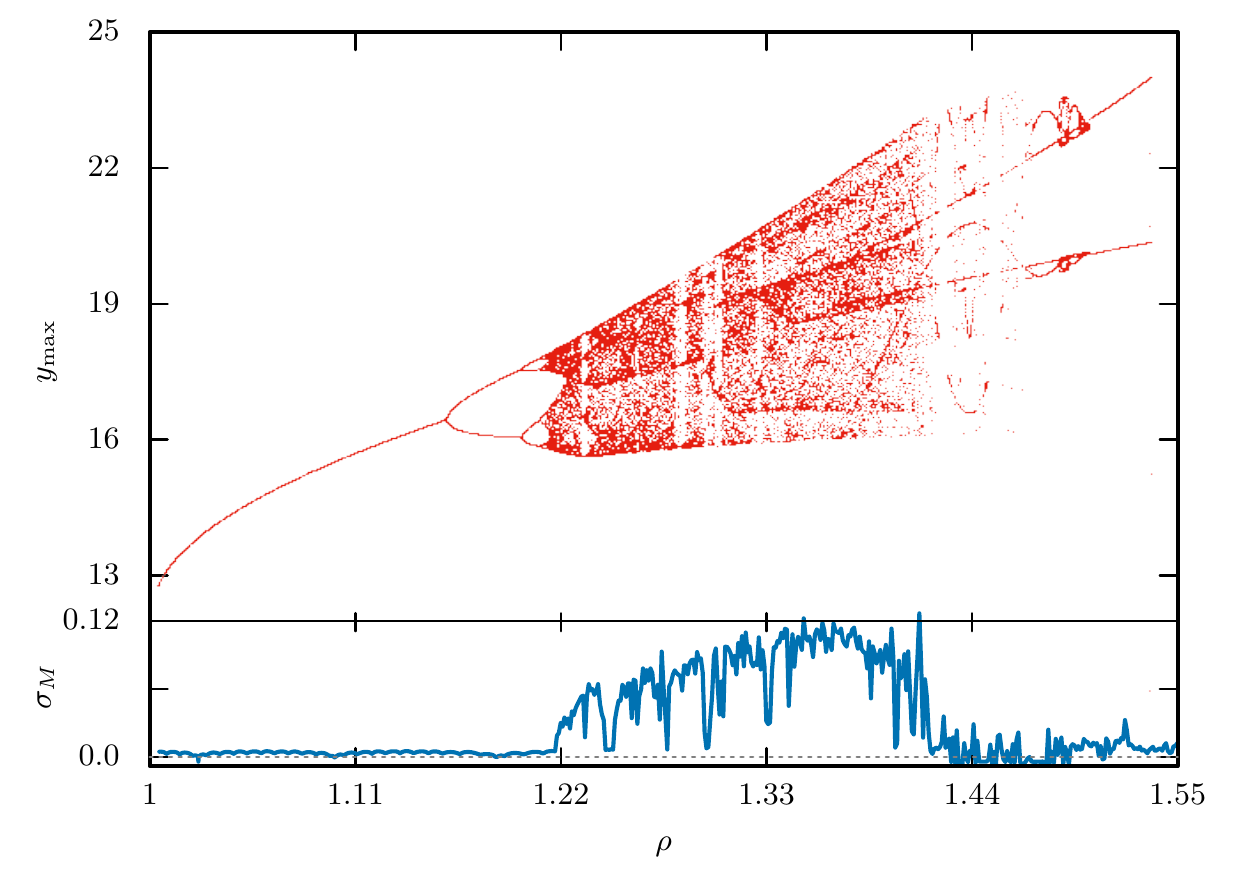}\hfill{}\includegraphics[width=0.5\textwidth]{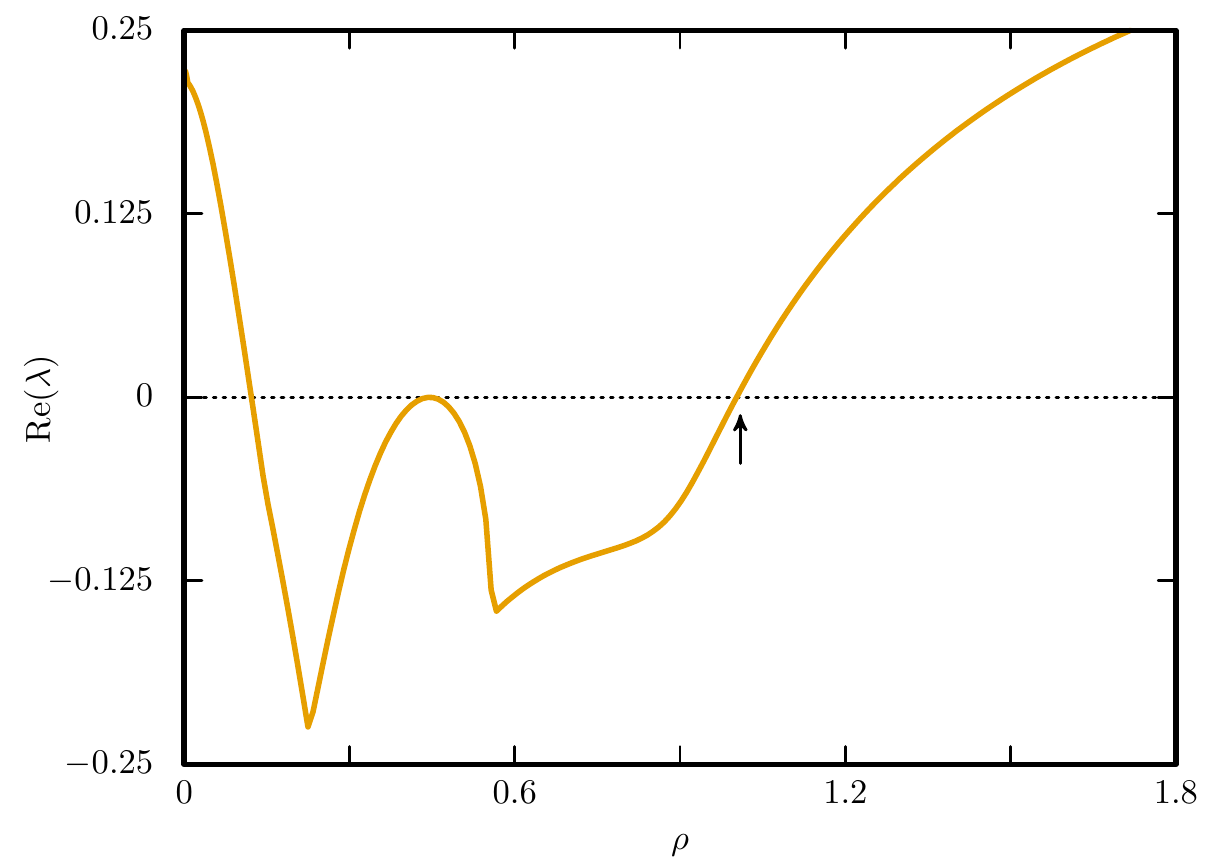}
\par\end{centering}
\caption{\label{fig:The-orbit-diagram}1st panel : The orbit diagram ($y_{{\rm max}}$
versus $\rho$) for the parameters $\beta=0.045,\eta=9.9,\tau_{1}=3.4,\tau_{2}=2.805$.
The corresponding maximal Lyapunov exponent $(\sigma_{M})$ is plotted
in the lower panel of the plot. 2nd panel : Plot of the real part
of the eigenvalue ${\rm Re(\lambda)}$ is shown against the free parameter
$\rho$. The arrow mark shows the position of the Hopf point.}
\end{figure}

As an example, we find that the parameter regime $\rho=1.53,\beta=0.045,\eta=9.9$
with $\tau_{1}=3.4$ and $\tau_{2}=2.805$ results a double peak pulse
as shown in the Fig.\ref{fig:The-phase-portrait}. Interestingly this
double-peak pulse is surrounded by chaotic oscillations on both sides.
To explore the bifurcation of Eqs.(\ref{eq:delay-1},\ref{eq:delay-2})
with these parameters, taking $\rho$ as the free parameter, we plot
the orbit diagram (a plot of maxima of $y$, $y_{{\rm max}}$ versus
$\rho$), starting from the Hopf point, along with the corresponding
maximal Lyapunov exponent $\sigma_{M}$ \cite{delay-technique} (first
panel of Fig.\ref{fig:The-orbit-diagram}). The Hopf point can be
found by solving Eqs.(\ref{eq:hopf1},\ref{eq:hopf2}) for the given
parameters for the variables $(\rho,\omega)$, which yields the solutions
$\rho\simeq1.00251,\omega\simeq0.69$. A plot of the real part of
the eigenvalue versus $\rho$, showing the Hopf point is also shown
in the second panel of Fig. \ref{fig:The-orbit-diagram}. As can be
seen from the orbit diagram that the first period doubling bifurcation
following the Hopf bifurcation occurs at $1.003$ and the subsequent
period doubling bifurcation occurs at $1.159$. After that the system
follows the usual route to chaos through a series of period doubling
bifurcations \cite{strogatz-book}, also characterised by the maximal
Lyapunov exponent. This chaotic regime continues till $\rho\simeq1.4$
after which a stable two-period pulse appears, which is our characteristic
pulse regime at low temperature. Beyond $\rho=1.55,$ the system exhibits
bouts of chaotic regimes and stable periodic oscillations akin to
logistic maps.

\section{Summary}

To summarise, we have constructed a low dimensional model for the
emission of firefly flash. We have compared our numerical findings
with the existing experimental results \cite{firfely-male}. Though
our model has certain limitations, it successfully explains the experimental
results describing the effect of temperature on amplitude and period
of the firefly flash. Our model is able to predict the rate constants
$k_{1}$ and $k_{2}$, which nearly agrees with the existing experimental
pyrosequencing data \cite{pyrosequencing}. Comparing our numerical
results with the experimental data \cite{firfely-male}, we conclude
that $k_{1}$ and $k_{2}$ both tend to increase with temperature
due to the rapidity of the production of the $\ce{Luc.LH}_{2}\textrm{-}\ce{AMP}$
inside the peroxisomes in quiescent phase of the lighting cycle and
in the flash phase inside the light emitting organ, $k_{2}$ is found
to be increases more rapidly compared to $k_{1}$, as it is the rate
constant for the production of light and as the production of $\ce{Luc.LH}_{2}\textrm{-}\ce{AMP}$
is larger in high temperature and oxygen is being continuously supplied
by the tracheoles, it obviously produces more light causing increase
in $k_{2}$, i.e. $\beta$ increases abruptly with temperature resulting
increase in the sharpness of the peak and decrease in the amplitude
of a pulse and the duration of a lighting cycle. 

Addition of time delay in our model explains low temperature behavior
of the firefly flash pattern. At low temperature \cite{firfefly-female,firfely-double-peak},
the firefly enzymatic catalytic reaction becomes slower and slower
with decreasing temperature resulting broadening of the peaks (distorted
into two or more than two peaks) and increase in the duration of the
lighting cycle. At lower temperature, $k_{1}$ for the production
of $\ce{Luc.LH}_{2}\textrm{-}\ce{AMP}$ at some earlier time is found
to be smaller compared to other non delayed case which obeying the
slowness of the reaction. Here, oxygen supplied in present time is
large enough compared to the past time that it combines with the delayed
$\ce{Luc.LH}_{2}\textrm{-}\ce{AMP}$ resulting increase in $k_{2}$
with a larger duration of flash with distorted peaks. Numerically,
formation of double peak at low temperature can be explained by the
period doubling bifurcation followed by a Hopf bifurcation. For the
delayed model, we produce a double peak following the trends of approximate
parameter values taken at non-delayed case. Maximal Lyapunov exponent
is further calculated for the delayed firefly model and positive Lyapunov
exponents implies the existence of chaotic regime. We have found lots
of local regime where multiple orbits are found and it is believed
to be controlled by firefly itself through the neurological responses.

\section*{Acknowledgement}

This work is carried out with a support from the SERB-DST (India)
research grant no. CRG/2018/002971. Necessary computational facilities,
in part, are provided through institutional FIST support (DST, India). 

\section*{Appendix}

\subsection{Approximation of a DDE system}

As we know that a dynamical system with time delay is basically an
\emph{infinite dimensional }system, the delay system can be reduced
to a set of $N$ ordinary differential equations, where $N$ is a
large positive integer \cite{delay-technique}.

Consider our DDE with two delays $\tau_{1,2}$,
\begin{eqnarray}
\frac{dx}{dt} & = & f[x(t),x(t-\tau_{1}),x(t-\tau_{2}),y(t)],\\
\frac{dy}{dt} & = & g[x(t),y(t)],
\end{eqnarray}
assuming $\tau_{1}>\tau_{2}$ numerically. We now divide the larger
of the delay into $N$ different discrete divisions, each of size
$\Delta t$ such that $\Delta t=\tau_{1}/N$ or equivalently $N=\tau_{1}/\Delta t$.
As $N\gg1$, $\Delta t$ is very small so that we have,
\begin{equation}
x(t+\Delta t)\simeq x(t)+\Delta t\,f[x(t),x(t-N\Delta t),x(t-M\Delta t),y(t)],
\end{equation}
where $M\,(<N)$ is an integer such that $M=\tau_{2}/\Delta t$ (please
see explanation about $M$ at the end of this section). Denoting now
the values of $x(t)$ at a discrete time $t_{j}$ as $x_{j}(t)$,
we have,
\begin{equation}
x_{j+1}(t)=x_{j}(t-\Delta t),
\end{equation}
which can be written as
\begin{eqnarray}
x_{j+N}(t) & = & x_{j}(t-N\Delta t),\nonumber \\
\Rightarrow x_{N}(t) & = & x_{0}(t-N\Delta t),
\end{eqnarray}
so that we can have now
\begin{equation}
\frac{d}{dt}x_{0}(t)=f[x_{0}(t),x_{N}(t),x_{M}(t),y(t)].
\end{equation}
Following this prescription, we can now write the original DDE system
as,
\begin{eqnarray}
\frac{dx_{0}}{dt} & = & f(x_{0},x_{N},x_{M},y),\\
\frac{dx_{j}}{dt} & = & \frac{N}{2\tau_{1}}(x_{j-1}-x_{j+1}),\quad1\leq j\leq N-1,\\
\frac{dx_{N}}{dt} & = & \frac{N}{\tau_{1}}(x_{N-1}-x_{N}),\\
\frac{dy}{dt} & = & g(x_{0},y),
\end{eqnarray}
for $N\gg1$, which can now be solved with any standard method. For
$N\to\infty$, the above system will reduce to the original DDEs.

One important point to be noted here about the value of $N$. While
any large positive integer as a value of $N$ will do for a single
delay DDE system, for a system with double delays such as in our case,
it is important that the difference between the two delays $(\tau_{1}-\tau_{2})$,
assuming $\tau_{1}>\tau_{2}$, is \emph{exactly} divisible by $\text{\ensuremath{\Delta t} }$
for the chosen value of $N$. This is required as we have now divided
the entire time interval $[0,\tau_{1}]$ into discrete divisions,
the smaller of the delays, which in this case is $\tau_{2}$ falls
exactly at one node, otherwise the discrete set of equations \emph{will
not} maintain the analytical difference between the delays for small
values of $N$. In our chosen set of values of $\tau_{1}=2.53$ and
$\tau_{2}=2.3$, $N$ should be a multiple of $11$. If the difference
between the delays is not \emph{exactly }divisible by $\Delta t$,
$N$ should be large enough to make the this difference equivalent
to the analytical value, otherwise even a small value of $N$ gives
quite good results.

\end{document}